\documentclass[fleqn,usenatbib]{mnras}

\usepackage{newtxtext,newtxmath}
\usepackage[T1]{fontenc}
\usepackage{ae,aecompl}

\usepackage{graphicx}	% Including figure files
\usepackage{amsmath}	% Advanced maths commands
\usepackage{amssymb}	% Extra maths symbols

\usepackage{subfig}

\usepackage{tikz}
\usetikzlibrary{bayesnet}

\usetikzlibrary{arrows,positioning, calc}
\tikzstyle{vertex}=[draw,fill=black!15,circle,minimum size=20pt,inner sep=0pt]

\usepackage{algorithm}
\usepackage[noend]{algpseudocode}

\algnewcommand\And{\textbf{and}}

\title[Streaming Classification of Variable Stars]{Streaming Classification of Variable Stars}

\author[L. Zorich et al.]{
L. Zorich,$^{1}$\thanks{E-mail: lezorich@uc.cl}
K. Pichara,$^{1,3,2}$\thanks{E-mail: kpb@ing.puc.cl}
P. Protopapas$^{2}$\thanks{E-mail: pavlos@seas.harvard.edu}
\\
$^{1}$Computer Science Department, Pontificia Universidad Cat\'olica de Chile, Santiago, Chile\\
$^{2}$Institute for Applied Computational Science, Harvard University, Cambridge, MA, USA\\
$^{3}$ Millenium Institute of Astrophysics, Santiago, Chile\\
}

\date{Accepted XXX. Received YYY; in original form ZZZ}

\pubyear{2018}

\begin{document}
\label{firstpage}
\pagerange{\pageref{firstpage}--\pageref{lastpage}}
\maketitle

\begin{abstract}
In the last years, automatic classification of variable stars has received substantial attention. Using machine learning techniques for this task has proven to be quite useful. Typically, machine learning classifiers used for this task require to have a fixed training set, and the training process is performed offline. Upcoming surveys such as the Large Synoptic Survey Telescope (LSST) will generate new observations daily, where an automatic classification system able to create alerts online will be mandatory. A system with those characteristics must be able to update itself incrementally. Unfortunately, after training, most machine learning classifiers do not support the inclusion of new observations in light curves, they need to re-train from scratch. Naively re-training from scratch is not an option in streaming settings, mainly because of the expensive pre-processing routines required to obtain a vector representation of light curves (features) each time we include new observations. In this work, we propose a streaming probabilistic classification model; it uses a set of newly designed features that work incrementally. With this model, we can have a machine learning classifier that updates itself in real time with new observations. To test our approach, we simulate a streaming scenario with light curves from CoRot, OGLE and MACHO catalogs. Results show that our model achieves high classification performance, staying an order of magnitude faster than traditional classification approaches.
\end{abstract}


%%%%%%%%%%%%%%%%%%%%%%%%%%%%%%%%%%%%%%%%%%%%%%%%%%

%%%%%%%%%%%%%%%%% BODY OF PAPER %%%%%%%%%%%%%%%%%%

\section{Introduction}

In the last years, automatic classification of variable stars has been heavily studied \citep{debosscher2007automated,richards2011,kim2011,bloom2011data,pichara2012,pichara2013,nun2014,huijse2014,nun2015fats,pichara2016,mackenzie2016clustering,benavente2017automatic,valenzuela2017automatic,cabreras-vives2017}. Machine learning techniques applied to photometric datasets has proven to be quite effective at this task. Commonly, these methods work by first extracting a vector of features from the light curve, which are statistical descriptors that represent different aspects of them; such as the light curve mean brightness, periodicity, and color, among others \citep{nun2015fats}.  After the feature extraction step, comes the learning stage ---or \textit{training}--- where machine learning classifiers learn patterns from the feature vectors having the classes ---or \textit{labels}---  of the objects. The initially labeled dataset is called \textit{training set}. Unfortunately, the whole process, the feature calculation, and the training stage are expensive in computing resources and typically are performed offline. Upcoming astronomical surveys will create an immense amount of data daily, making offline algorithms extremely costly. For example, the Large Synoptic Survey Telescope (LSST) \citep{borne2007}, which will start operating in 2022, will generate from 15 to 30TB of data per night. In fact, it is expected that it will produce the amount of data equivalent to scanning the whole available sky every three days. Having an updated classifier in such scenario (by retraining it on a frequent basis) to use it, for example, in early detection of alerts for brokers would be very resource-intensive. Today, automatic variable stars classification methods are not designed to work in a streaming setting, where new data is continuously arriving: for every new observation, algorithms that extract light curves features would need to reprocess the entire light curve again, and machine learning classifiers would need to go through the training process again in the whole dataset.

Most of machine learning approaches used in astronomy can produce high accuracy results \citep{big_universe}. However, these classification methods still work in an offline fashion. One of the most utilized sets of features used in variable stars classification are FATS features (Feature Analysis of Time Series) \citep{nun2015fats}, which contains more than 65 features available for the community in a Python library\footnote{\url{https://github.com/isadoranun/FATS}}. Using FATS in a streaming setting would mean a quadratic time complexity in the feature extraction stage, because every time new observations for a light curve arrives,all previous observations need to be processed again in order to update the light curve features with the new observations. On the other hand, for variable stars classification, the classifier of choice is usually the Random Forest \citep{random-forest}. Similarly, if we want to keep our classifier trained with the latest data, we would need to continuously train the Random Forest with the whole dataset every time new observations arrive. That is not a problem if the amount of data we are handling is small, but as we mentioned earlier, future surveys will generate terabytes of data daily.

The problem of how to efficiently maintain a classifier up to date with new data arises in many fields that deal with a continuous stream of information \citep{data_stream_survey}. In many areas of science and technology, massive streaming data is increasingly the norm \citep{big_data_1,big_data_2,big_data_3}. The machine learning community has proposed several streaming models to deal with this issue \citep{online-learning-review}. While approaches vary in nature, the purpose in most works is similar. \cite{mondrian_forests} uses Mondrian processes \citep{mondrian_process} to build an ensemble of random streaming decision trees, \citep{streaming-variational-bayes} proposes a framework for streaming, distributed and asynchronous Bayesian posterior inference, while \cite{bmm-gmm-jaini2016} uses Bayesian moment matching to update a Gaussian mixture model in an streaming fashion. Concerning streaming feature extraction algorithms for astronomical light curves, to the best of our knowledge little to no effort has been made.

In this work, we build on some of the features compiled in FATS, modifying them so their computation we can be updated in an streaming fashion. In addition, we present an incremental Bayesian Classification model, inspired by the work done in \cite{streaming-variational-bayes} and \cite{bmm-gmm-jaini2016}. Our Bayesian model can be efficiently updated, has a consistent treatment of uncertainty, and is generative, meaning that we can produce synthetic examples from it once it is trained.

The objective of this work is to introduce an end-to-end streaming method, from feature extraction to a probabilistic model, to classify variable stars in a streaming setting. The incremental light curve features proposed show competitive predictive performance comparable with traditional features used for variable stars classification, while being faster to compute in a normal setting, and an order of magnitude faster in a streaming setting. We also compared our approach with an streaming version of the Random Forest classifier \citet{mondrian_forests}, where our method shows an improvement in the results.

The reminder of this paper is organized as follows: Section~\ref{sec:related-work} gives an account of previous work on variable stars classification, light curve representation and streaming classification. Section~\ref{sec:background-theory} introduces the relevant background theory. Section~\ref{sec:methodology} describes the proposed method in detail. In section~\ref{sec:data}, the datasets used in this work are described. Section~\ref{sec:implementation} gives an account of the implementation details of the method, and in section~\ref{sec:results} we present our main results. Finally, in section~\ref{sec:conclusions}, we present the main conclusions of our work.

\section{Related Work}
\label{sec:related-work}

Most machine learning approaches used for automatic variable stars classification in astronomy work by first summarizing each light curve in a vector of statistical descriptors called features, and then training a classifier that learns patterns in these features \citep{bishop}. Many years of research effort has been applied in the development of the features for light curve representation. \cite{debosscher2007automated} proposed 28 features extracted from photometric analysis to represent each light curve. \cite{kim2009} used the Anderson-Darling test to test if a given light curve can be drawn from a Normal distribution. In later work, this test was added as a new feature. \cite{richards2011} introduced features derived from periodicity analysis using the Lomb-Scargle periodogram, as well as features like amplitude, standard deviation, kurtosis, skewness among others. \cite{kim2011} introduced features related to variability and dispersion. \cite{pichara2012} proposed using a continuous auto-regressive model to improve quasar detection. (Kim 2014) introduced features related to quartile analysis. \cite{nun2015fats} developed an open source library to help in the extraction of most features used in recent literature. \cite{mackenzie2016clustering} proposed an unsupervised method for automatically learning light curve features that work as good as traditional features.
However, to the best of our knowledge, most of the published work use offline algorithms for feature extraction, which does not match the streaming context.
	
After the feature extraction stage, machine learning models learn class-separation patterns relying on the feature vectors from labeled data. The machine learning community has proposed several classification models throughout the years: decision trees \citep{quinlan1993}, naive Bayes \citep{dudaHart1973}, Neural Networks \citep{rumelhart1986}, support vector machines \citep{cortes1995}, logistic regression \citep{cox1958} and Random Forest \citep{random-forest}. Gaussian Mixture models \citep{murphy2012} have also been used for classification, as well as other probabilistic models like Bayesian Networks \citep{friedman1997}, Latent Dirichlet Allocation \citep{blei2003}, and Gaussian processes \citep{rasmussen2005}, among others. Several of these models have been used to classify variable stars. \cite{debosscher2007automated} used a combination of Gaussian Mixture model, Support Vector Machines, Bayesian Networks and an artificial neural network to classify variable stars in OGLE \citep{ogle-III} and Hipparcos \citep{hipparcos}. \cite{richards2011} used a Random Forest, while \cite{pichara2013} combined it with Bayesian Networks to learn from catalogs with missing data. Also, in \citet{pichara2016} they presented a Meta-Classifier that learns how to re-use previously trained models to solve new variable stars classification scenarios without re-training from scratch.

Since in many areas of science and technology vast datasets of streaming data are increasingly the norm, lots of efforts have been made in the development of efficient streaming algorithms for this context, specially in the Bayesian paradigm.  \cite{hoffman2010} introduced streaming learning for Latent Dirichlet Allocation model using a streaming variational Bayes \citep{wainwright2008} algorithm, based on online stochastic optimization. \cite{streaming-variational-bayes} presented SDA-Bayes, a framework for making streaming updates to the Bayesian posterior in a distributed and asynchronous fashion. In the same fashion, \cite{campbell2015} introduced a framework for doing streaming and distributed inference in Bayesian nonparametric models. \cite{McInerney2015} proposed inferring a new type of posterior, the population posterior, which results from the application of Bayesian inference to a population distribution. \cite{bmm-gmm-jaini2016} introduced the Bayesian moment matching algorithm for learning a Gaussian mixture model in a streaming fashion. Similarly, \cite{hsu2016} also used streaming Bayesian moment matching algorithm for topic modeling with unknown number of topics in a streaming context.

To the best of our knowledge, the only work that tries to solve the problem of streaming classification of variable stars is \cite{klo2014}. Unfortunately, in their method they use traditional light curve features that are calculated in batch, and not incrementally. We view this as undesirable in a streaming context for variable stars classification, since when new observations arrive for a light curve, their features would need to be recalculated again. Given the immense amount of streaming data to be generated by future surveys, the development of an end-to-end efficient streaming classification pipeline for variable stars classification is a priority.
   
\section{Background Theory}
\label{sec:background-theory}

In this section, we briefly introduce the central concepts that are vital to understanding how our method works. First, we review the binary search tree data structure, which is key in the incremental feature extraction step of our method. Then we introduce the Gaussian mixture model, which is the primary building block of our streaming model. Finally, we explain why a Bayesian model naturally lends itself to a streaming setting, and how we can approximate the posterior of a Gaussian mixture model online using the Bayesian moment matching technique.

\subsection{Binary Search Tree}

A binary search tree \citep[]{Cormen2009} is a tree data structure in which each node has at most two children that satisfy the binary search tree property: all children in the left subtree of a node must hold a value smaller than its own, and all children in the right subtree of a node must hold a value larger than its own.

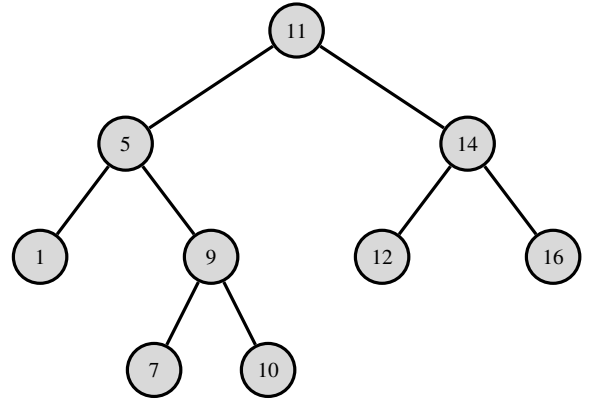
\begin{figure}
\begin{center}
\begin{tikzpicture}[very thick, level/.style={sibling distance=45mm/#1}]
\node [vertex] (r){$11$}
  child {
    node [vertex] (a) {$5$}
    child {
      node [vertex] {$1$}
    }
    child {
      node [vertex] {$9$}
      child {node [vertex] {$7$}}
      child {node [vertex] {$10$}}
    }
  }
  child {
    node [vertex] {$14$}
    child {
      node [vertex] {$12$}
    }
    child { 
      node [vertex] {$16$}
    }
  };
\end{tikzpicture}
\end{center}
\caption{The binary search tree data structure.}
\label{fig:binary-search-tree}
\end{figure}

For example, in Fig.~\ref{fig:binary-search-tree}, the value of the root node is $11$. The nodes with values $5$, $1$, $9$, $7$ and $10$ in its left subtree are no larger than $11$, and the nodes with values $14$, $12$ and $16$ in its right subtree are no smaller than $11$. This property is satisfied for every node in the tree, and it enables a fast lookup of a node in the tree. In fact, searching for a specific node is on average an $\mathcal{O}(\text{log}n)$ process. This is done by traversing the tree from root to leaf, comparing the value of the node we are searching for to the value of the current node to decide if we continue searching the left or right subtree. Similarly, inserting a new node also takes $\mathcal{O}(\text{log}n)$ operations on average.

\subsection{Gaussian Mixture Model}
\label{subsec:gmm_background}

\begin{figure}
  \includegraphics[width=\columnwidth]{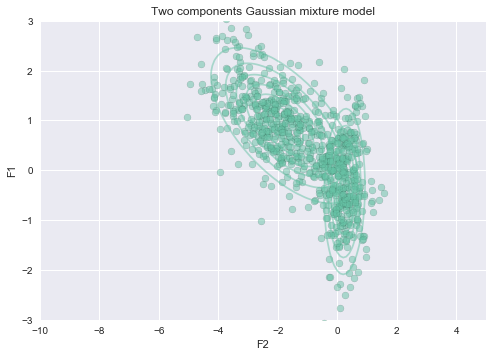}
  \caption{Example of 800 points drawn by the marginal distribution $P(\mathbf{x})$ of a mixture of 2 Gaussians. In this case, F1 and F2 can be any light curve feature.}
  \label{fig:two_comp_gmm}
\end{figure}

A Gaussian mixture model~\citep[]{murphy2012} is a linear combination of Gaussian distributions. In Figure \ref{fig:two_comp_gmm} shows an illustration of a mixture of two Gaussians. We call each Gaussian a \textit{component}. Let $K$ be the number of components in the mixture, and let $\mathbf{w}_i$ be 1-of-$K$ binary vector where $w_{ik}$ is equal to $1$ if $\mathbf{x}_i$ was generated by component $k$ and 0 otherwise. Let $\pi_k$ be the prior probability of a component assignment. We denote the observed data set as $\mathbf{X} = \{ \mathbf{x}_1, \dotsc, \mathbf{x}_N \}$ and the latent variables $\mathbf{Z} = \{ \mathbf{z}_1, \dotsc, \mathbf{z}_N \}$. With this, we can write the marginal distribution of $\mathbf{Z}$ as

\begin{equation}\label{eq:marginal-z-gmm}
	P(\mathbf{Z} \mid \boldsymbol{\pi}) = \prod_{i=1}^N\prod_{k=1}^K \pi_{k}^{z_{ik}}.
\end{equation}

Similarly, we can write the conditional distribution of $\mathbf{X}$ given the latent variables $\mathbf{Z}$ and the component parameters as
\begin{equation}\label{eq:conditional-x-gmm}
	P(\boldsymbol{X} \mid \mathbf{Z}, \boldsymbol{\mu}, \boldsymbol{\Lambda}) = \prod_{i=1}^N\prod_{k=1}^K\left[\mathcal{N}(x_i \mid \boldsymbol{\mu}_k, \boldsymbol{\Lambda}_k)\right]^{z_{ik}}
\end{equation}
where $\boldsymbol{\mu} = \{ \boldsymbol{\mu}_1, \dotsc, \boldsymbol{\mu}_K \}$,
$\boldsymbol{\Lambda} = \{ \boldsymbol{\Lambda}_1, \dotsc, \boldsymbol{\Lambda}_K \}$ 
and $\mathcal{N}(\mathbf{x}_i \mid\boldsymbol{\mu}_k, \boldsymbol{\Lambda}_k)$ is the probability density function for the multivariate Gaussian distribution with mean vector $\boldsymbol{\mu}_k$ and precision matrix $\boldsymbol{\Lambda}_k$. With equations (\ref{eq:marginal-z-gmm}) and (\ref{eq:conditional-x-gmm}) we can marginalize over $\mathbf{Z}$ to obtain the mixture's density function

\begin{equation}
	\begin{aligned}
		P(\boldsymbol{X} \mid \boldsymbol{\pi}, \boldsymbol{\mu}, \boldsymbol{\Lambda}) &= \prod_{i=1}^N\sum_{\mathbf{z}_i} P(\mathbf{x}_i \mid \mathbf{z}_i)P(\mathbf{z}_i) \\
        &= \prod_{i=1}^N\sum_{k=1}^K\pi_k\mathcal{N}(\mathbf{x}_i \mid \boldsymbol{\mu}_k, \boldsymbol{\Lambda}_k).
	\end{aligned}
\end{equation}

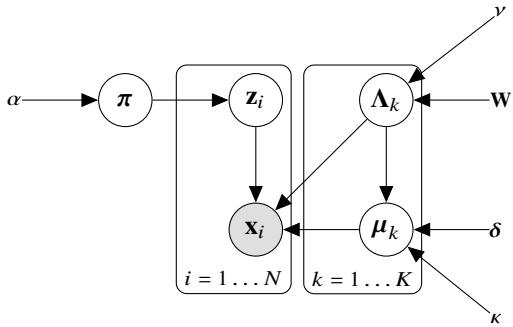
\begin{figure}
	\begin{tikzpicture}
      \node[obs]                   (x_n)      {$\mathbf{x}_i$} ; %
      \node[latent, above=of x_n]    (z_n)      {$\mathbf{z}_i$} ; %
      \node[latent, left=of z_n]    (pi)  {$\boldsymbol{\pi}$}; %
      \node[const, left=of pi]     (alpha) {$\alpha$};
      
      \node[latent, right=of x_n] (mu_k) {$\boldsymbol{\mu}_k$};
      \node[const, right=of mu_k] (gamma_0) {$\boldsymbol{\delta}$};
      \node[const, below=of gamma_0] (kappa) {$\kappa$};
      \node[latent, above=of mu_k] (lambda_k) {$\boldsymbol{\Lambda}_k$};
      \node[const, right=of lambda_k] (W) {$\mathbf{W}$};
      \node[const, above=of W] (nu) {$\nu$};
      
      \edge {z_n} {x_n};
      \edge {pi} {z_n};
      \edge {alpha} {pi};
      
      \edge {mu_k} {x_n}
      \edge {lambda_k} {mu_k};
      \edge {lambda_k} {x_n};
      \edge {W} {lambda_k};
      \edge {gamma_0} {mu_k};
      \edge {kappa} {mu_k};
      \edge {nu} {lambda_k};

      \platepgm {plate1} { %
        (x_n) %
        (z_n) %
      } {$i = 1 \dotsc N$}; %
      \platepgm {plate1} { %
        (lambda_k) %
        (mu_k) %	
      } {$k = 1 \dotsc K$}; %
    \end{tikzpicture}
    \caption{The plate notation for the Gaussian mixture model. The gray node denotes observed data. In this case, each data point $\mathbf{x}_i$ is observed and is generated by the component indicated by $\mathbf{z}_i$. There are K components, each of them with mean vector $\boldsymbol{\mu}_k$ and precision matrix $\boldsymbol{\Lambda}_k$. Nodes without a circle indicate prior hyperparameters.}
    \label{fig:gmm_plate_notation}
\end{figure}

Figure \ref{fig:gmm_plate_notation} shows the plate notation for the Gaussian mixture model. In our method, we use a mixture of Gaussians to model the light curve features.

\subsection{Streaming Bayesian inference and Bayesian Moment Matching for Gaussian mixture model}
\label{subsec:bmm-gmm}

Consider a stream of data $\mathbf{x}_1, \mathbf{x}_2, \ldots$ generated independent and identically distributed (\textit{iid}) by a distribution $p(\mathbf{x} \mid \Theta)$. In the context of our work, the data $\mathbf{x}_i$ corresponds to astronomical light curve features. Also, assume that $\Theta$ has the prior $p(\Theta)$ defined. Then, after given the first data point $\mathbf{x}_1$, Bayes theorem gives us the \textit{posterior distribution}

\begin{equation}
p(\Theta \mid \mathbf{x}_1) \propto p(\mathbf{x}_1 \mid \Theta)p(\Theta).
\end{equation}
When given the second data point $\mathbf{x}_2$, the posterior distribution becomes
\begin{equation}
\begin{split}
p(\Theta \mid \mathbf{x}_2, \mathbf{x}_1) &\propto p(\mathbf{x}_2, \mathbf{x}_1 \mid \Theta)p(\Theta) \\
&\propto p(\mathbf{x}_2 \mid \Theta)\underbrace{p(\mathbf{x}_1 \mid \Theta)p(\Theta)}_{\text{Posterior of }\mathbf{x}_1} \\
&\propto p(\mathbf{x}_2 \mid \Theta)p(\Theta \mid \mathbf{x}_1).
\end{split}
\end{equation}

As we can see, the streaming posterior $p(\Theta \mid \mathbf{x}_n, \mathbf{x}_{n-1}, \ldots, \mathbf{x}_1$, can be computed recursively using the posterior $p(\Theta \mid \mathbf{x}_{n-1}, \ldots, \mathbf{x}_1)$ as a prior. However, in our context, we need to estimate the streaming posterior of a Gaussian mixture model. In fact, if we expand $p(\Theta \mid \mathbf{x}_2, \mathbf{x}_1)$, we have that

\begin{equation}
p(\Theta \mid \mathbf{x}_2, \mathbf{x}_1) 
	\propto p(\boldsymbol{\pi}, \boldsymbol{\mu}, \boldsymbol{\Lambda} \mid \mathbf{x}_1)\sum_{k=1}^K \pi_k\mathcal{N}(\mathbf{x}_1 \mid \boldsymbol{\mu}_k, \boldsymbol{\Lambda}_k^{-1}).
\end{equation}
As a result, if we compute the posterior $p(\Theta \mid \mathbf{x}_N, \mathbf{x}_{N-1}, \ldots \mathbf{x}_1)$ recursively, because of the summation over the number of components ($K$), the number of terms in the posterior grows exponentially. In fact, there will be $K^N$ terms in the posterior after $N$ data points, which is intractable. To circumvent this problem, the Bayesian moment matching (BMM) algorithm for Gaussian mixture model was proposed by ~\cite{bmm-gmm-jaini2016, bmm-psm-jaini2016}. BMM approximates the exact posterior $p(\Theta \mid \mathbf{x}_N, \mathbf{x}_{N-1}, \ldots, \mathbf{x}_1)$ by a distribution $\tilde{p}(\Theta)$ by matching a set of sufficient moments. The advantage of BMM is that it lends itself naturally online, thus the Gaussian mixture model posterior can be updated after each data point $\mathbf{x}$ of the stream arrives. If we choose a Dirichlet prior over the weights $\boldsymbol{\pi}$ and a Normal-Wishart prior over the parameters $(\boldsymbol{\mu}, \boldsymbol{\Lambda}^{-1})$, the posterior distribution after the first data point $\mathbf{x}_1$ arrives is

\begin{equation}
	\begin{split}
		p(\Theta \mid \mathbf{x}_1) 
        &\propto p(\Theta)\sum_{k=1}^K \pi_k\mathcal{N}(\mathbf{x}_1 \mid \boldsymbol{\mu}_k, \boldsymbol{\Lambda}_k^{-1}) \\
        &= Dir(\boldsymbol{\pi} \mid \boldsymbol{\alpha}) \prod_{j = 1}^K \mathcal{NW}(\boldsymbol{\mu}_j, \boldsymbol{\Lambda}_j \mid \boldsymbol{\delta}, \kappa, \mathbf{W}, \nu) \\
        &\qquad\qquad\qquad\sum_{k=1}^K \pi_k\mathcal{N}(\mathbf{x}_1 \mid \boldsymbol{\mu}_k, \boldsymbol{\Lambda}_k^{-1}).
	\end{split}
\end{equation}

\begin{figure*}
  \includegraphics[width=\textwidth]{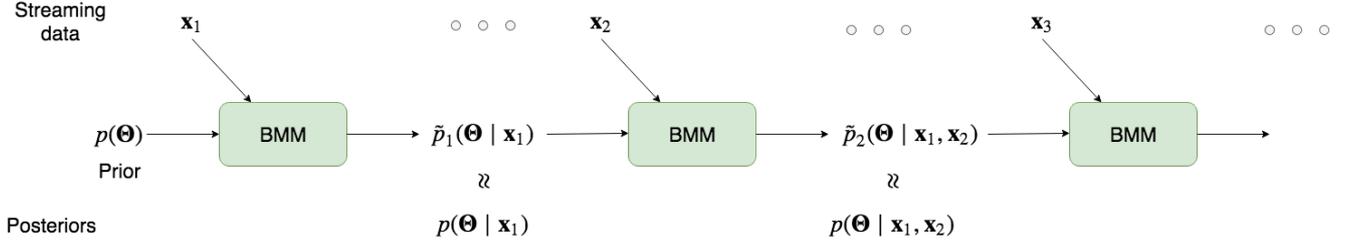}
  \caption{Illustration of the Bayesian moment matching algorithm.}
  \label{fig:streaming_bmm}
\end{figure*}

Since the Dirichlet and Normal-Wishart distribution are conjugate priors for the likelihood, the posterior $p(\boldsymbol{\Theta} \mid \mathbf{x}_1)$ has the same form as the family of distributions of the prior $p(\boldsymbol{\Theta})$. Then, the Bayesian moment matching method approximates the posterior $p(\boldsymbol{\Theta} \mid \mathbf{x}_1)$ by a distribution $\tilde{p}(\boldsymbol{\Theta} \mid \mathbf{x}_1)$, which is of the same family of the prior $p(\boldsymbol{\Theta})$ by matching all the sufficients moments of $p(\boldsymbol{\Theta} \mid \mathbf{x}_1)$ with $\tilde{p}(\boldsymbol{\Theta} \mid \mathbf{x}_1)$. The approximate posterior $\tilde{p}(\boldsymbol{\Theta} \mid \mathbf{x}_1)$ contains a single product of Dirichlet and $K$ Normal-Wishart distributions

\begin{equation}	
	\tilde{p}_1(\boldsymbol{\Theta} \mid \mathbf{x}_1) = Dir(\boldsymbol{\pi} \mid \boldsymbol{\alpha}^1)\prod_{j=1}^K \mathcal{NW}(\boldsymbol{\mu}_j, \boldsymbol{\Lambda}_j \mid \boldsymbol{\delta}^1, \kappa^1, \mathbf{W}^1, \nu^1).
\end{equation}
The parameters $\boldsymbol{\gamma}_0^1, \boldsymbol{\delta}^1, \kappa^1, \mathbf{W}^1, \nu^1$ are evaluated by matching the set of sufficient moments of $p(\boldsymbol{\Theta} \mid \mathbf{x}_1)$, $S = \{ \boldsymbol{\mu}_j, \boldsymbol{\mu}_j\boldsymbol{\mu}_j^T, \boldsymbol{\Lambda}_j, \boldsymbol{\Lambda}^2_{jkm}, \pi_j, \pi_j^2 \}$ for $j = 1, \ldots, K$, with $\tilde{p}(\boldsymbol{\Theta} \mid \mathbf{x}_1)$, where $\boldsymbol{\Lambda}_{jkm}$ is the $(k, m)$ element of $\boldsymbol{\Lambda}_j$. Figure \ref{fig:streaming_bmm} shows an illustration of the BMM algorithm. For an more in depth explanation of the method, see \cite{bmm-gmm-jaini2016}. 

\section{Methodology}
\label{sec:methodology}

\begin{figure}
  \includegraphics[width=\columnwidth]{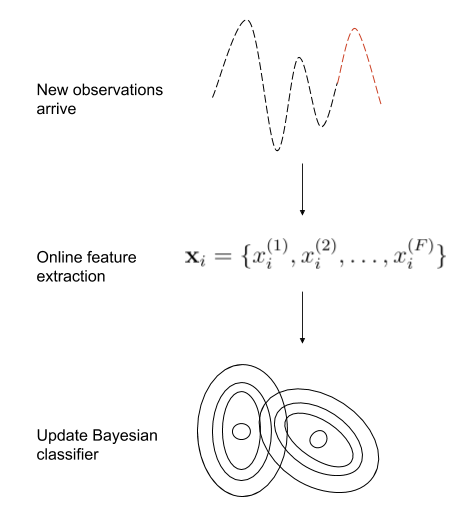}
  \caption{The proposed method process.}
  \label{fig:main_process}
\end{figure}

Our method consists of two steps: the streaming feature extraction step, and the streaming classifier. We use a subset of the features compiled in \cite{nun2015fats} and re-design them to work in a streaming fashion. This results in some of the features being an approximation of the offline ones. The second stage consists of a streaming Bayesian classifier based on Gaussian Mixture models, explained in Section \ref{subsec:gmm_background}. We use the Bayesian Moment Matching algorithm, explained in Section \ref{subsec:bmm-gmm}, to update the classifier incrementally. The method process is illustrated in Figure \ref{fig:main_process}.

\subsection{Incremental feature extraction}
\label{subsec:incremental_feature_extraction}

The first step of our method consists of the incremental light curves feature extraction.
Because this is a streaming process, we don't know the real number of observations the light curve has.
From now on, $N$ is the current number of observations we have for a light curve. 

\subsubsection{Mean}
\label{subsubsec:mean}

The light curve mean magnitude is calculated using the equation

\begin{equation}
\label{eq:mean-mag}
	\bar{m} = \frac{\sum_{i=1}^N m_i}{N}.
\end{equation}
If $\bar{m}_{N}$ is the mean magnitude when the first $N$ observations have been seen, then, when a new observation $m_{N+1}$ arrives, the new magnitude is calculated as

\begin{equation}
\label{eq:mean-mag}
	\bar{m}_{N+1} = \frac{N\bar{m}_{N} + m_{N + 1}}{N + 1}.
\end{equation}

\subsubsection{Standard deviation}

The sample standard deviation is defined as

\begin{equation}
\label{eq:std}
	\sigma = \sqrt[]{\frac{1}{N - 1}\sum_{i=1}^N (m_i - \bar{m})^2}.
\end{equation}
If $\sigma_{N}$ is the standard deviation calculated with the first $N$ observations, then, we can approximate the standard deviation when observation $m_{N+1}$ arrives as

\begin{equation}
\label{eq:std}
	\sigma_{N+1} \approx \sqrt{\frac{1}{N} \left[\sigma_{N}^2 (N - 1) + (m_{N + 1} - \bar{m}_{N + 1})^2\right]}.
\end{equation}

\subsubsection{Mean variance}

The mean variance is a variability index and can be calculated as

\begin{equation}
	\text{mv} = \frac{\bar{m}}{\sigma}.
\end{equation}
When a new observation arrives, the new approximate mean variance is updated using the new mean $\bar{m}_{N+1}$ and standard deviation $\sigma_{N+1}$.

\subsubsection{Range of a cumulative sum}

The range of cumulative sum \citep{cumsum} of a lightcurve is defined as

\begin{equation}
	\text{R}_{\text{cs}} = \text{max}S - \text{min}S
\end{equation}
where,

\begin{equation}
	\text{S}_l = \frac{1}{N\sigma} \sum_{i=1}^l (m_i - \bar{m}) = \frac{1}{N\sigma} (-l\bar{m} + \sum_{i=1}^l m_i) = \frac{1}{N\sigma}(-l\bar{m} + w_l)
\end{equation}
where $w_l = \sum_{i=1}^l m_i$ for $l = 1, 2, \dotsc, N$.\\

For incrementally update this feature, we need to keep track of the maximum $\text{S}$ and the minimum $\text{S}$. With this intention, for every $l = 1, 2, \dotsc, N$ we keep the sum $w_l = \sum_{i=1}^l m_i$ and the index $l$ as node in a binary search tree using $w_l$ as key. When the new observation $m_{N+1}$ arrives, we add a new node to the binary search tree with key $w_{l + 1} = w_l + m_{N+1}$ and index $l+1$. Obtaining the new $\text{min}S$ and $\text{max}S$, is identically to obtaining the minimum and maximum $-(l+1)\bar{m}_{N+1} + w_{l+1}$ term, because $\frac{1}{(N + 1)\sigma_{N+1}}$ remains constant for all $l = 1, 2, \dotsc, N + 1$. Then, we can follow a greedy search algorithm \citep{Cormen2009} using the binary search tree. We start from the root of the tree, and compare the value of $-(l+1)\bar{m}_{N+1} + w_{l+1}$ with the left child and the right child. When we want to find the minimum, if the root has the smallest value, then return the value of the root; in the case that the left child or the right child has the smallest value, set that child as the root and repeat the process; and lastly, if the tree has no more children, then the minimum has been found. For finding the maximum, the process is similar, but instead of guiding the search to the smallest value, we use the largest value. The algorithm for finding the minimum is shown in Algorithm \ref{alg:cumsum}.

\begin{algorithm}
\caption{Gets the minimum of $-(l+1)\bar{m}_{N+1} + w_{l+1}$, which corresponds to the value that minimizes $S$. \textit{root} refers to the current tree node, \textit{root}.left is the left child of the \textit{root} node and  \textit{root}.right is the right child of the \textit{root} node. The algorithm first checks if the current node is a leaf node. If the current node is not a leaf node, then it searches for the smallest value in its left or right child.}\label{alg:cumsum}
\begin{algorithmic}[1]
\Procedure{GetMin}{$root, \bar{m}_{N+1}$}
\If {$root\text{.left} = \text{NIL} \And root\text{.right} = \text{NIL}$}
\State \textbf{return} root
\EndIf
\State $rootsum \gets -root\text{.l}\cdot\bar{m}_{N+1} + root.w$
\If {$root\text{.left} \not= \text{NIL}$}
\State $leftsum \gets -root\text{.left.l}\cdot\bar{m}_{N+1} + root\text{.left.w}$
\If {$leftsum < rootsum$}
\State \textbf{return} \textsc{GetMin}(root\text{.left}, $\bar{m}_{N+1}$)
\EndIf
\EndIf
\If {$root\text{.right} \not= \text{NIL}$}
\State $righttsum \gets -root\text{.right.l}\cdot\bar{m}_{N+1} + root\text{.right.w}$
\If {$rightsum < rootsum$}
\State \textbf{return} \textsc{GetMin}(root.right, $\bar{m}_{N+1}$)
\EndIf
\EndIf
\State \textbf{return} root
\EndProcedure
\end{algorithmic}
\end{algorithm}

Once we find the minimum and maximum, we subtract the minimum to the maximum to obtain the updated $\text{R}_{\text{cs}}$. The whole process takes $\mathcal{O}(\text{log} n)$ time. Although the algorithm is not fully online, it is fast enough for our purposes.

\subsubsection{Period (Analysis of Variance)}

In \cite{kim2011} the Lomb-Scargle \citep{scargle} algorithm for period estimation is proposed. The problem is that the time complexity of the fast algorithm is $\mathcal{O}(n \text{log}(n))$ \citep{fast-lomb-scargle}. With the intention of incrementally update the period of the light curve, we propose using the analysis of variance \citep{aov1996} as a period search method.

This method tests the significance of a trial period using a standard statistical method in the light curve folded and grouped into bins. The two statistics used are

\begin{equation}
s_1 = \frac{1}{r - 1}\sum_{i=1}^r b_i (\bar{m}_i - \bar{m})^2
\end{equation}

\begin{equation}
s_2 = \frac{1}{N - r}\sum_{i=1}^r \sum_{j=1}^{b_i} (m_{ij} - \bar{m}_i)^2
\end{equation}
where $r$ is the number of bins, $b_i$ is the number of observations in the $i$th bin, $\bar{m}_i$ is the average of the observations in the $i$th bin and $m_{ij}$ is the $j$th observation of the $i$th bin. As a result, the period of the light curve is the trial period with the maximum $\frac{s_1}{s_2}$.

In the case of $s_1$, when a new observation arrives, we need to determine to which of the bins it corresponds, and update $b_i$, $\bar{m}_i$ and $\bar{m}$ accordingly. The variables $\bar{m}_i$ and $\bar{m}$ can be updated using the formula shown in Section \ref{subsubsec:mean}.

On the other hand, $s_2$ can be rewritten as

\begin{equation}
s_2 = \frac{1}{N - r}\sum_{i=1}^r \left[ b_i\bar{m}_i^2 -2\bar{m}_i\left(\sum_{j=1}^{b_i} m_{ij} \right) + \left(\sum_{j=1}^{b_i} m_{ij}^2 \right) \right].
\end{equation}
Then, if we maintain the sum $\sum_{j=1}^{b_i} m_{ij}$ and $\sum_{j=1}^{b_i} m_{ij}^2$, we can calculate $s_2$ in an incremental way.

As a result, the time complexity of updating the light curve period when a new observation arrives is constant with respect to the number of observations of the light curve. It only depends on the number of bins and the number of periods we try.

\subsubsection{Color}

The light curve color~\citep{kim2011} is the difference 
between the mean magnitude of two different bands. When a new observation arrives, we update the mean magnitude of each band and then take the difference between them.

\subsubsection{Stetson K}
\label{subsubsec:stetson_k}

Stetson K, Stetson J and Stetson L features \citep{kim2011}, which are in Section \ref{subsubsec:stetson_j} and \ref{subsubsec:stetson_l} respectively, are based on the Welch/Stetson variability index I \citep{stetson}. In particular, Stetson K is a a robust kurtosis measure, and is defined as

\begin{equation}
K = \frac{N^{-1} \sum_{i=1}^N \mid \delta_i \mid}{\sqrt{N^{-1} \sum_{i=1}^N \delta_i^2}}
\end{equation}
where $\delta_i$ is the relative error, and is defined as

\begin{equation}
\delta_i = \sqrt{\frac{N}{N - 1}} \frac{m_i - \hat{m}_N}{\epsilon_i}
\end{equation}
where $\epsilon_i$ is the standard error of the magnitude $m_i$. 

If we maintain the sum $\psi^{(1)}_N = \sum_{i=1}^N \mid \frac{m_i - \hat{m}_{N}}{\epsilon_i} \mid$ and $\psi^{(2)}_N = \left( \frac{m_i - \hat{m}_N}{\epsilon_i} \right)^2$, when a new observation arrives, we can update $\sum_{i=1}^{N + 1} \mid \delta_i \mid$ as

\begin{equation}
\sum_{i=1}^{N+1} \mid \delta_i \mid = \sqrt{\frac{N + 1}{N}} \left(\psi^{(1)}_N + \frac{m_{N+1} - \hat{m}_{N+1}}{\epsilon_{N+1}} \right).
\end{equation}
Similarly, we can update $\sum_{i=1}^N \delta_i^2$ as

\begin{equation}
\sum_{i=1}^{N+1} \delta_i^2 = \frac{N+1}{N} \left( \psi^{(2)}_{N+1} + \frac{m_{N+1} - \hat{m}_{N+1}}{\epsilon_{N+1}} \right).
\end{equation}

With the updates of $\sum_{i=1}^{N + 1} \mid \delta_i \mid$ and $\sum_{i=1}^N \delta_i^2$, we can update Stetson K of the light curve in a streaming manner.

\subsubsection{Stetson J}
\label{subsubsec:stetson_j}

Stetson J is calculated using two bands of the same light curve and is defined as

\begin{equation}
J_N = \sum_{i=1}^N sgn(\text{P}_i)\sqrt{\mid \text{P}_i \mid}
\end{equation}
where $\text{P}_i$ is the product between $\delta_i$, defined in Section \ref{subsubsec:stetson_k}, of the two different bands. That is, $\text{P}_i = \delta^{(a)}_i\delta^{(b)}_i$.

Because we can calculate $\delta_i$ incrementally, when a new observation arrives we can update $J$ with the following equation

\begin{equation}
J_{N+1} = J_N + sgn(\text{P}_{N+1})\sqrt{\mid \text{P}_{N+1} \mid}.
\end{equation}

\subsubsection{Stetson L}
\label{subsubsec:stetson_l}

Stetson L is defined as

\begin{equation}
L = \frac{JK}{0.798}.
\end{equation}
It represents the synchronous variability of different bands. Since we already know how to update $J$ and $K$ in a streaming manner, updating $L$ is straightforward.

\subsubsection{Flux percentile ratio mid 20, mid 35, mid 50, mid 65 and
	mid 80}

If we define $F_{p_1,p_2}$ as the difference between $p_2\%$ and $p_1\%$ magnitude values, we calculate the flux percentile ratio features \citep{richards2011} as:

\begin{itemize}
\item{Flux percentile ratio mid 20: $F_{40,60}/F_{5,95}$}
\item{Flux percentile ratio mid 35: $F_{32.5,67.5}/F_{5,95}$}
\item{Flux percentile ratio mid 50: $F_{25,75}/F_{5,95}$}
\item{Flux percentile ratio mid 65: $F_{17.5,82.5}/F_{5,95}$}
\item{Flux percentile ratio mid 80: $F_{10,90}/F_{5,95}$}
\end{itemize}

In order to incrementally update these features, we use a binary search tree to keep the light curve magnitudes sorted. When a new observation arrives, we insert it in a binary search tree. Then, we use Algorithm \ref{alg:flux} to get the element at position $\lfloor (N + 1) \frac{p}{100} \rceil$, where $p$ is the percentile value we want. The runtime complexity of updating the flux percentile ratio feature takes $\mathcal{O}(lg N)$ time.

\begin{algorithm}
\caption{Gets element at position $\lfloor (N + 1) \frac{p}{100} \rceil$. \textit{root} refers to the current tree node, \textit{root}.left is the left child of the \textit{root} node and  \textit{root}.right is the right child of the \textit{root} node. The algorithm checks if the element $k$ is in the left or right subtree, depending in the number of elements in each subtree. If $k$ is in neither subtree, then it returns the current node.}\label{alg:flux}
\begin{algorithmic}[1]
\Procedure{GetElementAtPosition}{$root, k$}
\If {$k < num\_elements(root\text{.left})$}
\State \textbf{return} \textsc{GetElementAtPosition}($root\text{.left}, k$)
\EndIf
\If {$k > num\_elements(root\text{.left})$}
\State $k \gets k - num\_elements(root\text{.left})$ 
\State \textbf{return} \textsc{GetElementAtPosition}($root\text{.right}, k$)
\EndIf
\State \textbf{return} root
\EndProcedure
\end{algorithmic}
\end{algorithm}

\begin{table}
	\centering
	\caption{Time complexity of the incremental features.}%
	\label{tab:features}
    \begin{tabular}{ l l r }
        \hline
        Feature & Exact & Time complexity \\
        \hline
        Color & Yes & $\mathcal{O}(1)$ \\
        StetsonJ & No & $\mathcal{O}(1)$ \\
        StetsonL & No & $\mathcal{O}(1)$ \\
        Period & Yes & $\mathcal{O}(1)$ \\
        RangeCS & No & $\mathcal{O}(\text{log} N)$ \\
        StetsonK & No & $\mathcal{O}(1)$ \\
        FluxPercentileRatioMid50 & Yes & $\mathcal{O}(\text{log} N)$ \\
        FluxPercentileRatioMid65 & Yes & $\mathcal{O}(\text{log} N)$ \\
        Mean & Yes & $\mathcal{O}(1)$ \\
        FluxPercentileRatioMid35 & Yes & $\mathcal{O}(\text{log} N)$ \\
        FluxPercentileRatioMid20 & Yes & $\mathcal{O}(\text{log} N)$ \\
        FluxPercentileRatioMid80 & Yes & $\mathcal{O}(\text{log} N)$ \\
        Mean Variance & No & $\mathcal{O}(\text{log} N)$ \\
        Std & No & $\mathcal{O}(1)$ \\
        \hline
    \end{tabular}
\end{table}

\subsection{Classification model}

Consider a stream of light curve features $\mathbf{x}_i$, for $i = 1, 2, \ldots$ where $\mathbf{x}_i = \{ x_i^{(1)}, x_i^{(2)}, \ldots, x_i^{(F)} \}$ with $F$ as the number of features that are going to be used for classification. This features can come from either a new light curve, or a light curve which their features were updated because new observations arrived. Also, consider that the light curve features come with an already known label. We represent the class of $\mathbf{x}_i$ as a 1-of-M representation, $\mathbf{z}_i$, where $z_{im} = 1$ if $\mathbf{x}_i$ is of class $m$, and $z_{im} = 0$ otherwise. This is similar to what was defined in \ref{subsec:gmm_background}, but in this case $\mathbf{x}_i$ represents the light curve features, and $z_{i}$ represents the light curve class.

\begin{figure}
  \includegraphics[width=\columnwidth]{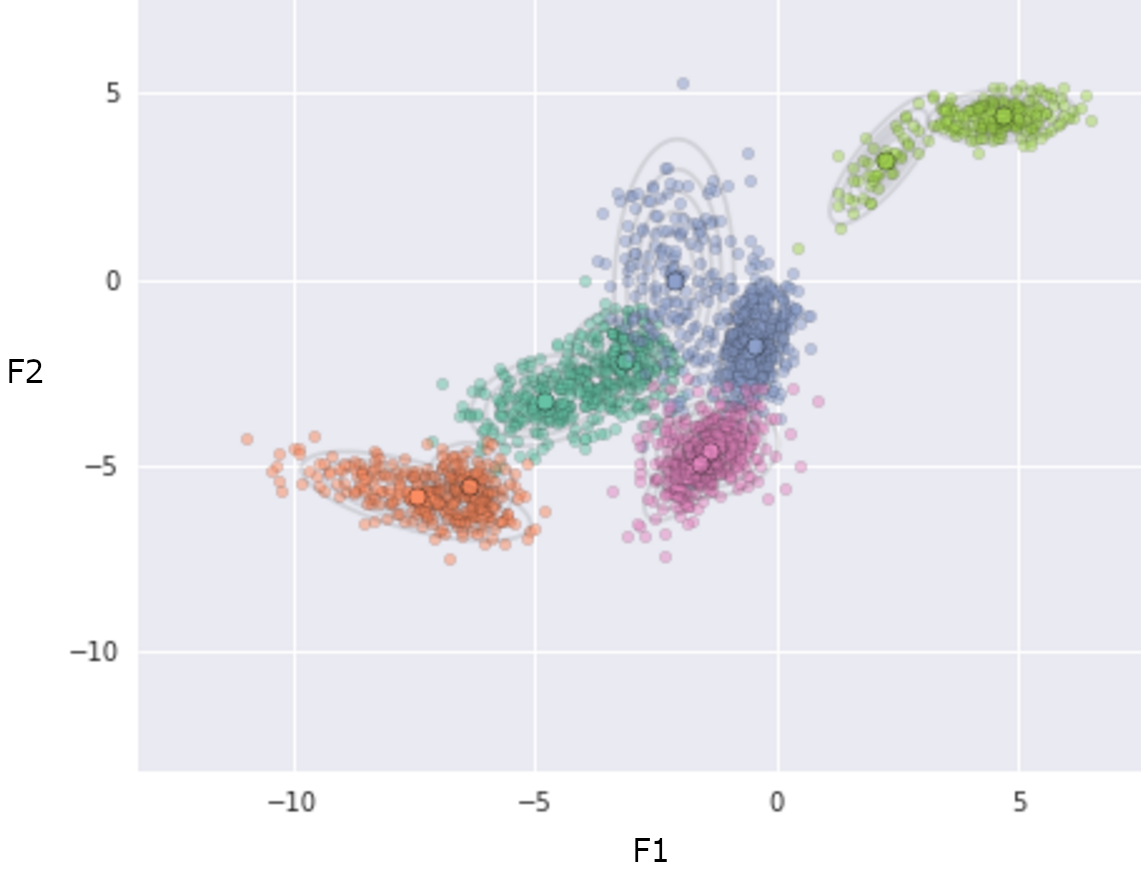}
  \caption{Example of five different classes (one class per color), where each class is modeled as a Gaussian mixture model. The axis can be any light curve feature.}
  \label{fig:mgmm}
\end{figure}

Provided that it is well known that a finite mixture of Gaussian densities can approximate arbitrarily any continuous distribution \citep{lindsay95, MCLA2000}, we use a mixture of Gaussians to model the data in the feature space. We use one Gaussian mixture model per class, with $K$ components each mixture. In other words, there are $M$ Gaussian mixtures, with $M$ the number of light curve variability classes. As an example, in the Figure \ref{fig:mgmm} we can see 5 different classes where each one is modeled with a Gaussian mixture of two components, thus in the case of the Figure, $M = 5$.

Then, given that $\mathbf{x}_i$ is of class $m$, and that it was generated by the component $k$ of the Gaussian mixture of class $m$, we have that

\begin{equation}
\text{P}(\mathbf{x}_i \mid \mathbf{z}_i = m, \mathbf{w}_{im} = k, \boldsymbol{\mu}, \mathbf{\Lambda}) = \mathcal{N}(\mathbf{x}_i \mid \boldsymbol{\mu}_{mk}, \mathbf{\Lambda}_{mk}^{-1})
\end{equation}
where $\boldsymbol{\mu} = \{\{\boldsymbol{\mu}_{mk}\}\}$ and $\boldsymbol{\Lambda} = \{\{\boldsymbol{\Lambda}_{mk}\}\}$, for $m = 1, \dotsc, M$ and $k = 1, \dotsc, K$. The random variables $\boldsymbol{\mu}_{mk}$ and $\boldsymbol{\Lambda}_{mk}$ are the mean and precision matrix for the $k$-th component from the $m$-th Gaussian mixture. We use a Normal-Wishart prior for $\boldsymbol{\mu}_{mk}$ and $\boldsymbol{\Lambda}_{mk}$

\begin{equation}
\text{P}(\boldsymbol{\mu}_{mk}, \boldsymbol{\Lambda}_{mk}) = \mathcal{NW}(\boldsymbol{\mu}_{mk}, \boldsymbol{\Lambda}_{mk} \mid \boldsymbol{\mu}_0, \beta_0, \mathbf{W}_0, \nu_0).
\end{equation}
The responsibility\footnote{Responsibility refers to the estimation of the total amount of data points generated by a particular Gaussian component.} of each Gaussian component of class $m$ is defined as

\begin{equation}
\text{P}(\mathbf{w}_{im} \mid \boldsymbol{\rho}_m) = \text{Cat}(\mathbf{w}_{im} \mid \boldsymbol{\rho}_m).
\end{equation}
The random variable $\mathbf{w}_{im}$ is a 1-of-K vector which represents from which of the $k$ components $\mathbf{x}_i$ of class $m$ was generated from. It is modeled as a categorical variable $\text{P}(\mathbf{w}_{im} \mid \boldsymbol{\rho}_m) = \text{Cat}(\mathbf{w}_{im} \mid \boldsymbol{\rho}_m)$, where $\mathbf{\rho}_{m}$ are the weights of the different components of the mixture, and it has a Dirichlet prior $\text{P}(\boldsymbol{\rho}_m \mid \boldsymbol{\gamma}_0) = \text{Dir}(\boldsymbol{\rho}_m \mid \boldsymbol{\gamma}_0)$.

Similarly, the light curve class $\mathbf{z}_i$ also is generated from a categorical distribution $\text{P}(\mathbf{z}_i \mid \boldsymbol{\pi}) = \text{Cat}(\mathbf{z}_i \mid \boldsymbol{\pi})$ with a Dirichlet prior $\text{P}(\boldsymbol{\pi} \mid \boldsymbol{\alpha}_0) = \text{Dir}(\boldsymbol{\pi} \mid \boldsymbol{\alpha}_0)$. 

\begin{figure}
	\begin{tikzpicture}
      \node[obs]                   (x_i)      {$\mathbf{x}_i$} ; %
      \node[obs, left=of x_i]    (z_i)      {$\mathbf{z}_i$} ; %
      \node[latent, left=of z_i]    (pi)  {$\boldsymbol{\pi}$}; %
      \node[const, above=of pi]     (alpha_0) {$\boldsymbol{\alpha}_0$};
      
      \node[latent, above=of z_i] (w_im) {$\mathbf{w}_{im}$};
      \node[latent, above=of  w_im] (rho_m) {$\boldsymbol{\rho}_m$};
      \node[const, above=of rho_m] (gamma_0) {$\boldsymbol{\gamma}_0$};
      
      \node[latent, right=of rho_m] (mu_mk) {$\boldsymbol{\mu}_{mk}$};
      \node[latent, right=of mu_mk] (lambda_mk) {$\boldsymbol{\Lambda}_{mk}$};
      \node[const, right=of lambda_mk] (w_0) {$\mathbf{W}_{0}$};
      \node[const, above=of lambda_mk] (v_0) {$\nu_{0}$};
      \node[const, above=of mu_mk] (mu_0) {$\boldsymbol{\mu}_{0}$};
      \node[const, left=of mu_0] (beta_0) {$\beta_{0}$};
      
      \edge {z_i} {x_i};
      \edge {pi} {z_i};
      \edge {alpha_0} {pi};
      
      \edge {w_im} {x_i};
      \edge {rho_m} {w_im};
      \edge {gamma_0} {rho_m};
      
      \edge {mu_mk} {x_i}
      \edge {lambda_mk} {mu_mk};
      \edge {lambda_mk} {x_i};
      \edge {w_0} {lambda_mk};
      \edge {v_0} {lambda_mk};
      \edge {mu_0} {mu_mk};
      \edge {beta_0} {mu_mk};

      \platepgm {plate1} { %
        (x_i) %
        (z_i) %
        (w_im) %
      } {$i = 1 \dotsc N$}; %
      \platepgm {plate2} { %
      	(w_im) %
        (rho_m) %
        (lambda_mk) %
        (mu_mk) %	
      } {$m = 1 \dotsc M$}; %
      \platepgm {plate3} { %
        (lambda_mk) %
        (mu_mk) %
      } {$k = 1 \dotsc K$}; %
    \end{tikzpicture}
    \caption{Plate notation for the proposed model. Gray nodes denote observed data. In this case, each data point $\mathbf{x}_i$ is observed and is generated by the Gaussian mixture indicated by the observed class $\mathbf{z}_i$. There are $M$ Gaussian mixtures, one per class. Each of them has  $K$ components with mean vector $\boldsymbol{\mu}_k$ and precision matrix $\boldsymbol{\Lambda}_k$. Nodes without a circle indicate prior hyperparameters.}
    \label{fig:mgmm_plate_notation}
\end{figure}
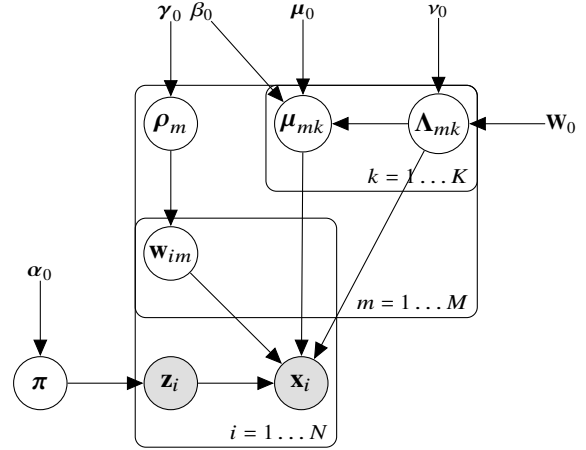

The graphical model can be seen in Figure \ref{fig:mgmm_plate_notation}.

\subsubsection{Streaming model inference}

Given the stream $\mathbf{x}_1, \mathbf{x}_2, \ldots$ and $\mathbf{z}_1, \mathbf{z}_2, \ldots$, we are interested in inferring the posterior distribution of the model in a streaming manner, where the posterior is given by

\begin{equation}\label{eq:posterior}
\text{P}(\boldsymbol{\pi}, \boldsymbol{\rho}, \boldsymbol{\mu}, \boldsymbol{\Lambda} \mid \mathbf{X}, \mathbf{Z}) \propto \text{P}(\mathbf{X}, \mathbf{Z} \mid \boldsymbol{\pi}, \boldsymbol{\rho}, \boldsymbol{\mu}, \boldsymbol{\Lambda})\text{P}(\boldsymbol{\pi}, \boldsymbol{\rho}, \boldsymbol{\mu}, \boldsymbol{\Lambda})
\end{equation}
where $\mathbf{X} = \{ \mathbf{x}_1, \mathbf{x}_2, \ldots \}$, $\mathbf{Z} = \{ \mathbf{z}_1, \mathbf{z}_2, \ldots \}$, $\boldsymbol{\rho} = \{ \boldsymbol{\rho}_m \}$, for $m = 1, \ldots, M$ and $\text{P}(\mathbf{X}, \mathbf{Z} \mid \boldsymbol{\pi}, \boldsymbol{\rho}, \boldsymbol{\mu}, \boldsymbol{\Lambda})$ is the model likelihood defined by

\begin{equation}
\text{P}(\mathbf{X}, \mathbf{Z} \mid \boldsymbol{\pi}, \boldsymbol{\rho}, \boldsymbol{\mu}, \boldsymbol{\Lambda}) = \prod_{i=1}^N \prod_{m=1}^M \left[ \sum_{k=1}^K \rho_{mk}\text{P}(\mathbf{x}_i \mid \boldsymbol{\mu}_{mk}, \boldsymbol{\Lambda}_{mk}^{-1}) \right]^{z_{im}}.
\end{equation}
Because the class $\mathbf{z}_i$ is observed, the latent variable $\boldsymbol{\pi}$ is d-separated \citep{dseparation} from the rest of the model given $\mathbf{z}_i$. Hence, we can separate the inference of $\text{P}(\boldsymbol{\pi} \mid \mathbf{Z})$, from the inference of $\text{P}(\boldsymbol{\rho}, \boldsymbol{\mu}, \boldsymbol{\Lambda} \mid \mathbf{X}, \mathbf{Z})$.

Since the Dirichlet distribution is the conjugate prior for the categorical distribution \citep{conjugate_prior_dir}, the exact posterior of $\text{P}(\boldsymbol{\pi} \mid \mathbf{Z})$ can be obtained analytically and is given by

\begin{equation}
\text{P}(\boldsymbol{\pi} \mid \textbf{Z}) = \text{Dir}(\mathbf{C} + \boldsymbol{\alpha}_0)
\end{equation}
where $\mathbf{C} = (c_1, \ldots, c_M)$, $c_m = \sum_i^N I(z_{im} = 1)$ and $I(z_{im} = 1)$ is the indicator function that has the value $1$ if $\mathbf{z}_i$ is of class $m$, and $0$ otherwise. In the case a feature of a new light curve arrives, we can obtain the new posterior by updating the variable $\mathbf{C}$ depending on the class $\mathbf{z}_{N+1}$. On the other hand, since $\mathbf{x}_i$ is restricted to have been generated by the Gaussian mixture of class $m$, we can infer the posterior of each Gaussian mixture independently. We make use of the Bayesian Moment Matching algorithm \citep{bmm-gmm-jaini2016}, explained in Section \ref{subsec:bmm-gmm}, in order to infer the posterior of each Gaussian mixture in a streaming setting.

\subsubsection{Classification}

We can use the Bayes theorem to find the probability that a light curve with features $\textbf{x}^{\star}$ belong to the variability class $m$

\begin{equation}
\begin{split}
\text{P}(\textbf{z}^{\star} = m \mid \textbf{x}^{\star}) &= \frac{\text{P}(\textbf{z}^{\star} = m, \textbf{x}^{\star})}{\text{P}(\textbf{x}^{\star})} \\
&= \frac{\text{P}(\textbf{x}^{\star} \mid \textbf{z}^{\star} = m)\text{P}(\textbf{z}^{\star} = m)}{\sum_{i} \text{P}(\textbf{x}^{\star} \mid \textbf{z}_i)\text{P}(\textbf{z}_i)} \\
&= \frac{\pi_m \left[ \sum_{k=1}^K \rho_{mk} \mathcal{N}(\textbf{x}^{\star} \mid \boldsymbol{\mu}_{mk}, \boldsymbol{\Lambda}_{mk}^{-1}) \right]}{\sum_{j=1}^M \pi_m \left[\sum_{k=1}^K \rho_{jk} \mathcal{N}(\textbf{x}^{\star} \mid \boldsymbol{\mu}_{jk}, \boldsymbol{\Lambda}_{jk}^{-1}) \right]}.
\end{split}
\end{equation}
Hence, the light curve is assigned to the class with highest posterior probability. 

\section{Data}
\label{sec:data}
For our experiments we use photometric data from three different catalogs: MACHO, OGLE-III and CoRoT.

\subsection{MACHO Catalog}

The Massive Compact Halo Object (MACHO) \citep{macho} project is a survey which started on 1992 and ended on 1999. Its primary objective is to detect microlensing events produced by the Galactic Bulge, the Large Magellanic Cloud (LMC) and the Small Magellanic Cloud (SMC). Light curves were obtained in the blue and red band. For our experiments, we use a set of 2,092 previously labeled light curves. The class distribution of the data is shown Table~\ref{tab:macho_class_dist}.

\begin{table}
	\centering
	\caption{Class distribution of MACHO labeled set.}
	\label{tab:macho_class_dist}
    \begin{tabular}{ l r }
        \hline
        Class Name & No. of elements \\
        \hline
        Quasar & 59 \\
        Be Star & 127 \\
        Cepheid & 101 \\
        RR Lyrae & 609 \\
        Eclipsing Binary & 254 \\
        MicroLensing & 578 \\
        Long Period Variable & 364 \\
        \hline
    \end{tabular}
\end{table}

\subsection{OGLE-III Catalog of Variable Stars}

The Optical Gravitational Lensing Experiment (OGLE) \citep{ogle} started in 1992 with the objective of finding microlensing events on the Magellanic Clouds and Galactic Bulge. In 2001, the third phase, OGLE-III Catalog of Variable Stars (OIII-CSV) \citep{ogle-III}, began, collecting a significant amount of labeled data.

We use $414,360$ labeled light curves from OIII-CSV in our experiments. Despite that the photometric data produced by this survey are in two bands, the I-band and the V-band, light curves in the V-band do not have enough observations for our purpose. Hence, for our experiments, we only use the I-band. The class distribution of the data is shown in Table~\ref{tab:ogle_class_dist}.

%Paragraph 1: Give a small summary of  what OGLE-III is.

%Paragraph 2: Talk about the labeled objects used in our experiments.

\begin{table}
	\centering
	\caption{Class distribution of OGLE-III labeled set.}
	\label{tab:ogle_class_dist}
    \begin{tabular}{ l r }
        \hline
        Class Name & No. of elements \\
        \hline
        Cepheid & $8,610$ \\
        RR Lyrae & $44,218$ \\
        Eclipsing Binary & $17,739$ \\
        Long Period Variable & $343,786$ \\
        \hline
    \end{tabular}
\end{table}

\subsection{CoRoT Catalog}

The Convection, Rotation, and Transit (CoRot) \citep{corot1,corot2} satellite was launched in 2006 with the objective of searching for exoplanets using transit photometry. It continuously observes the milky way for periods up to 6 months with a cadence that can be more than 100 observations per object per day. These observations are in the red, green and blue bands.

For our experiments, we use 1,311 labeled light curves from CoRoT. We combine the red, green and blue bands to form a white band, and use this four our experiments. Table \ref{tab:corot_class_dist} shows the class distribution for this dataset.

\begin{table}
	\centering
	\caption{Class distribution of CoRoT labeled set.}
	\label{tab:corot_class_dist}
    \begin{tabular}{ l r }
        \hline
        Class Name & No. of elements \\
        \hline
        Cepheid & $125$ \\
        RR Lyrae & $509$ \\
        Eclipsing Binary & $109$ \\
        Long Period Variable & $568$ \\
        \hline
    \end{tabular}
\end{table} 

\section{Implementation}
\label{sec:implementation}

Our method is implemented in Python 3.5 using numpy~\citep{numpy} for efficient numerical computation, pandas~\citep{pandas} for data manipulation and scipy~\citep{scipy} for probability distributions. We also use cython~\citep{cython} for an efficient AOV periodogram implementation. The code is available at \url{https://github.com/lezorich/variational-gmm}.

The parameters used in our experiments are detailed in Table~\ref{tab:parameters}. The number of Gaussians $K$ per mixture is optimized using cross-validation on the MACHO training set.

\begin{table}
	\centering
	\caption{Values of relevant parameters used for our experiments.}
	\label{tab:parameters}
    \begin{tabular}{ l l r }
        \hline
        Parameter & Meaning & Value \\
        \hline
        K & Number of Gaussians per mixture & 2 \\
        B & Number of light curve observations per batch & 20 \\
        \hline
    \end{tabular}
\end{table} 

\section{Experimental Results}
\label{sec:results}

In this section, we present the results from the application of the proposed method on the two catalogs described in Section~\ref{sec:data}. First, in Section~\ref{subsec:incremental_feat_results}, we present the results of the feature extraction step of our method, where we show that in a streaming setting our method is considerably more scalable than FATS features \citep{nun2015fats}. Then, in Section~\ref{subsec:classification_results}, we present the classification results on the three training sets described before using three different classifiers: the standard Random Forest \citep{random-forest}, the streaming Mondrian Forest \citep{mondrian_forests} and our approach. The intention is showing the difference in classification performance when using our incremental features instead of traditional FATS, and using a streaming model against conventional non-incremental classifiers.

\subsection{Analysis of Incremental features}
\label{subsec:incremental_feat_results}

\begin{figure}
  \includegraphics[width=\columnwidth]{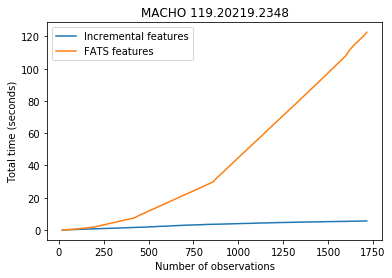}
  \caption{The time it takes to calculate the features shown in Table~\ref{tab:features} with different number of observations. We can see in the figure that the running time grows exponentially when using FATS library, while when using our method the running time grows linearly.}
  \label{fig:fats_vs_incremental_time}
\end{figure}

Given that the incremental features are proposed as a replacement for non-incremental FATS features in the streaming setting, one would expect that: the time to calculate them grows linearly with the number of observations and that they achieve comparable classification results as FATS. The latter will be shown in Section~\ref{subsec:classification_results}. To show the results of the former, in Figure~\ref{fig:fats_vs_incremental_time} we show the cumulated time it takes to calculate the features shown in  Table~\ref{tab:features} with our method and FATS. We can see in that Figure that the time it takes to calculate FATS grows quadratically with the number of observations. That is because when new observations arrive, FATS features are re-calculate with all observations again. Hence, FATS in a streaming setting takes $\mathcal{O}(N^2)$ time, where $N$ is the number of observations in the light curve. On the other hand, our proposed features only take $\mathcal{O}(N)$ to be processed.

The effects of the difference in time between both methods can be better seen when calculating the features in more than one light curve. In Table~\ref{tab:runtime_feature_extraction} we see the approximated computational runtime when extracting the features in MACHO, OGLE-III and CoRoT datasets in three different ways: first, using our incremental features (IF); second, using FATS  in a standard setting (no streaming) with the entire light curve (FATS-B); lastly, using FATS simulating a streaming scenario )FATS-I). All execution times are calculated using 32 cores. As we can see in that Table, our method is significantly faster than using FATS, including the case when using FATS in a no streaming setting. It is important to note that when the number of light curves increases, the difference in times is more noticeable. For instance, in the case of OGLE-III, we could not even extract the features in a streaming way using FATS because the time it took was excessively long.

\begin{figure*}
\label{feature-results}
\begin{tabular}{ccc}
\subfloat{\includegraphics[width = 2in]{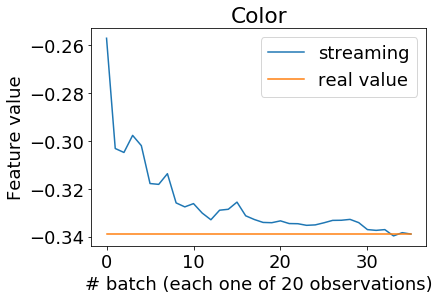}} &
\subfloat{\includegraphics[width = 2in]{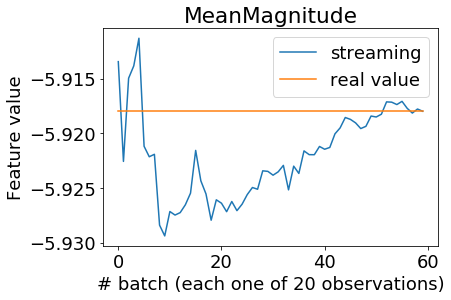}} &
\subfloat{\includegraphics[width = 2in]{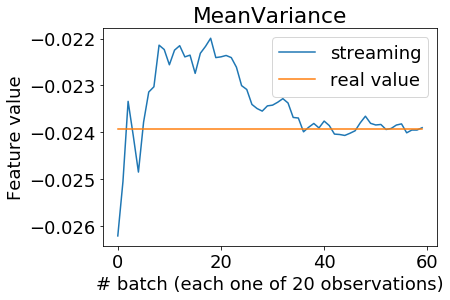}} \\
\subfloat{\includegraphics[width = 2in]{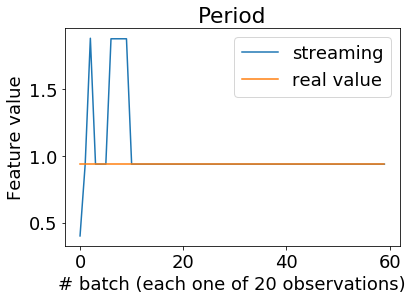}} &
\subfloat{\includegraphics[width = 2in]{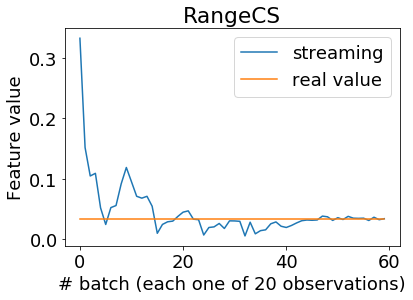}} &
\subfloat{\includegraphics[width = 2in]{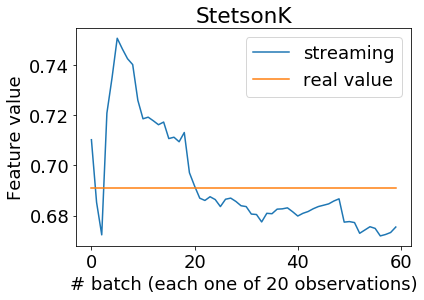}} \\ 
\subfloat{\includegraphics[width = 2in]{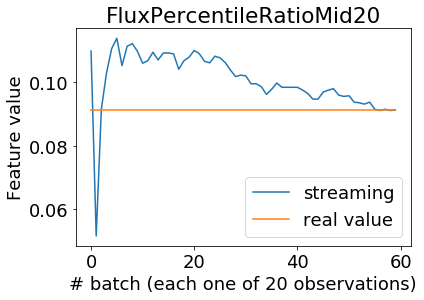}} &
\subfloat{\includegraphics[width = 2in]{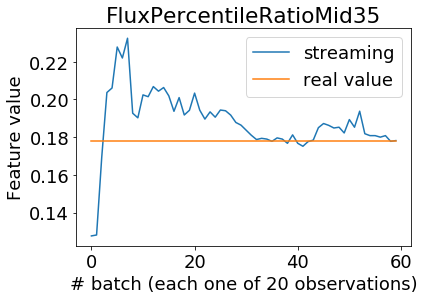}} &
\subfloat{\includegraphics[width = 2in]{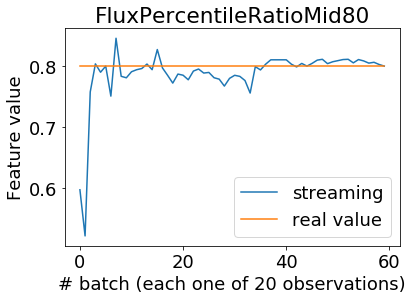}} \\
\subfloat{\includegraphics[width = 2in]{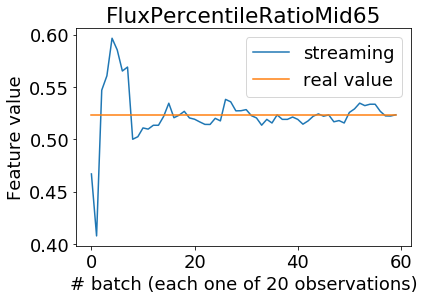}} &
\subfloat{\includegraphics[width = 2in]{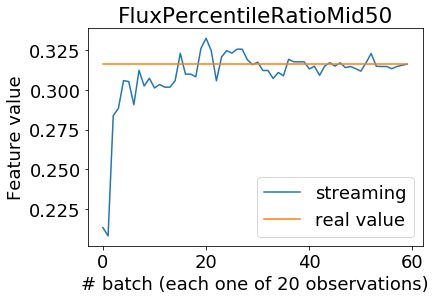}} &
\subfloat{\includegraphics[width = 2in]{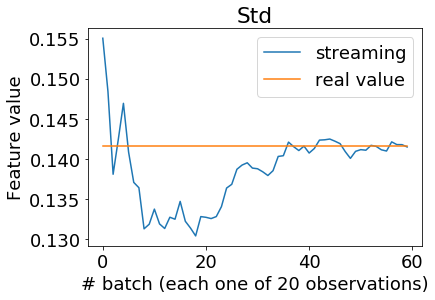}} \\
\subfloat{\includegraphics[width = 2in]{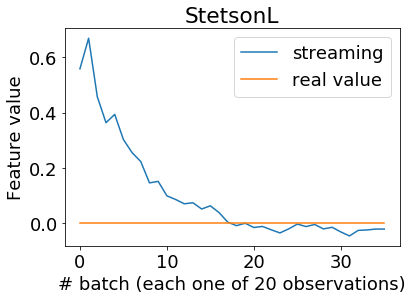}} &
\subfloat{\includegraphics[width = 2in]{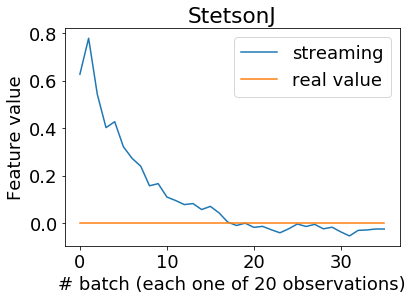}} &
\end{tabular}
\caption{Change of the value of the different streaming features for MACHO 1.3444.614 when new observations arrive. From the figure, we can appreciate that most features get close to their real value with few observations.}
\label{fig:fats_vs_incremental_value}
\end{figure*}

\begin{figure*}
 \includegraphics[width=\textwidth,]{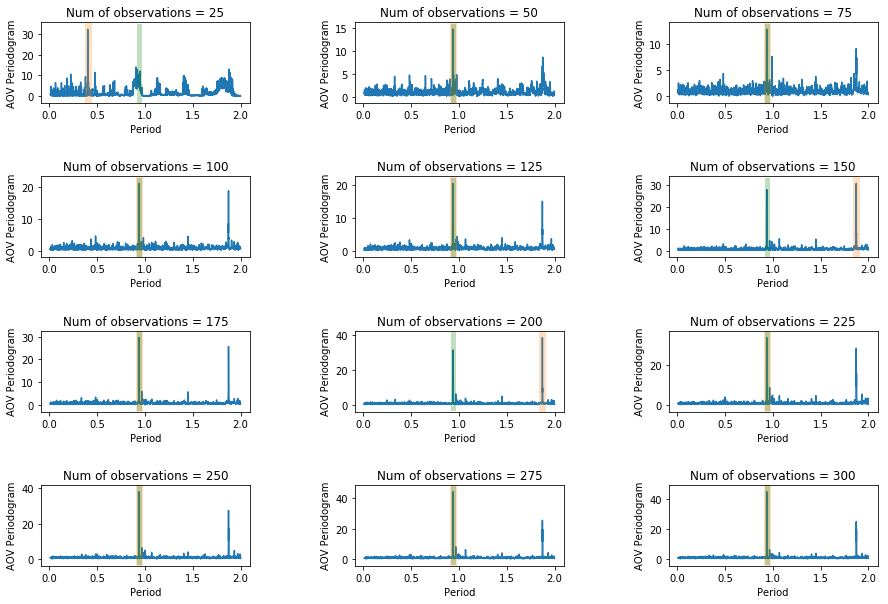}
 \caption{Change in the periodogram of an Eclipsing Binary as new observations arrive. The true period is shaded with a green line while the maximum of the periodogram is shaded with an orange line. We can appreciate that the real period is found with around 225 observations, that is around $19\%$ of this star total observations.}
 \label{fig:aov_periodogram}
\end{figure*}

Another result worthy of mention is that most incremental features converge, or get very close, to the real value of the feature (the one calculated offline with the whole light curve), with a relatively low number of observations. In Figure~\ref{fig:fats_vs_incremental_value} shows the value of all the incremental features while new observations arrive (in batches of 20), each for a light curve of the MACHO dataset. As we can see from the figure, with just 200 observations, almost all features get very close to the ground-truth value. The period, for example, converges entirely in the 10th batch. In Figure \ref{fig:aov_periodogram} we can see how the periodogram of a Eclipsing-Binary changes as new observations appear. With this previous fact, one would expect that the classification results using the incremental features should converge with a low number of observations too. As we will see in the following section, that is indeed what happens.

%Show the "error" between the incremental features and the standard FATS. The idea is to show that the incremental features are good approximation of the FATS features.

% Mostrar "diferencia entre features incrementales" y los otros features
% Incremental: 5.85 minutes en total

\begin{table}
	\centering
	\caption{Computational runtime details for feature extraction.}
	\label{tab:runtime_feature_extraction}
    \begin{tabular}{ l l l l }
        \hline
         & MACHO & OGLE-III & CoRoT \\
        \hline
        IF & \textbf{5 minutes} & \textbf{12 hours} & \textbf{2 minutes} \\
        FATS-B & 18 minutes & 3 days & 7 minutes \\
        FATS-I & 72 minutes & - & 28 minutes \\
        \hline
    \end{tabular}
\end{table}

\subsection{Classification results}
\label{subsec:classification_results}

To perform an initial test to check the correctness of our incremental approach, we simulate data in a streaming scenario. First, for each dataset, we randomly select in a stratified fashion $10\%$ of the light curves for the test set and left the rest as the training set. For the training set, we randomly select a set of $L_1$ light curves and train an initial model with them having their first $20$ observations. Then, we simulate that $20$ new observations arrive for a subset of $L_2$ light curves of the initial set of $L_1$ light curves. We call this a new batch. We simulate that several of those batches arrive and we update our model with each of those batches. We measure the classifier performance using F-Score with 10-fold\footnote{We randomly partition the original sample in 10 equal size subsamples, used 1 subsample as the test set and the remaining 9 as the training set. We repeat this process 10 times with each of the 10 subsamples (10-fold).} stratified by class cross-validation on each class on each training set and average the results to obtain the final F-Score. We use F-Score as the classification metric because it is a balance between both exactness (precision) and completeness (recall) and it is better suited than other metrics for imbalanced data like in this case. To classify light curve data by using time series features, the classifier of choice is usually the Random Forest \citep{richards2011,pichara2012}. Hence, we compare our streaming Bayesian classifier against a traditional Random Forest and a streaming Random Forest (Mondrian Forest) \citep{mondrian_forests} using the incremental features. For our streaming classifier and the streaming Random Forest, we choose $L_1$ as $1000$ in MACHO dataset, as $100000$ in the OGLE dataset and $400$ in CoRoT dataset. In addition, we choose $L_2$ as $100$, $1000$ and $100$ respectively. Because the traditional Random Forest does not work with streaming data, instead of training the model with the light curves selected for that batch, we retrain the model with the entire dataset again. Additionally, we also compare the proposed method against a Random Forest trained on all FATS with the intention of showing how much accuracy in classification is lost using our method.

\begin{figure}
  \includegraphics[width=\columnwidth]{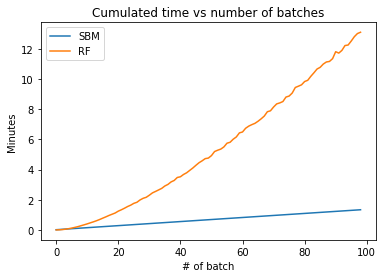}
  \caption{The evolution of the F-Score in OGLE as new observations arrives.}
  \label{fig:cum_time_classifier}
\end{figure}

Tables~\ref{tab:macho_classification_results}, \ref{tab:ogle_classification_results} and \ref{tab:corot_classification_results} show the F-Score achieved for each training set. The acronym IF refers to incremental features, SBM refers to our streaming Bayesian classifier, RF refers to the Random Forest and oRF refers to the streaming Random Forest. Results show that although being only $14$ features versus the $47$ features of FATS, and with some of the proposed features being an approximation of the real feature, incremental features are a good alternative when looking for scalability when doing variable stars classification. We can see the F-Score obtained with FATS+RF as the base line to have an intuition of the F-Score we get with the incremental (approximated) methods. Concerning the proposed SBM, it performs close to the RF, but the RF has better F-Score in general. It must be remembered that the RF is entirely retrained with every batch, while our SBM is updated online with new data only. As a result, it is expected that the SBM loses accuracy while being much more scalable. However, we can see that our model performs better than the streaming Random Forest in almost all classes. Also, Figure \ref{fig:cum_time_classifier} shows the cumulated time of training both models with a different number of batches in the OGLE-III dataset. In every batch shown in the Figure, a fixed amount of $30$ light curves changed. As we can see, when the number of batches grows, the cumulated time of training the RF grows more than the cumulated time of training the SBM. That is because the SBM is updated with just the $30$ light curves with new observations, while the RF needs to be retrained with the whole dataset again.

\begin{figure}
  \includegraphics[width=\columnwidth]{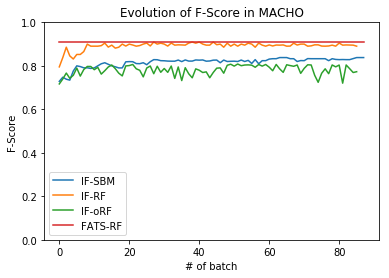}
  \caption{The evolution of the F-Score in MACHO as new observations arrives.}
  \label{fig:f_score_macho}
\end{figure}

\begin{figure}
  \includegraphics[width=\columnwidth]{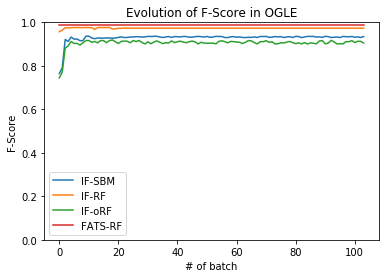}
  \caption{The evolution of the F-Score in OGLE as new observations arrives.}
  \label{fig:f_score_ogle}
\end{figure}

Figure \ref{fig:f_score_macho} and \ref{fig:f_score_ogle} shows the evolution of the F-Score while new batches arrive. In the case of OGLE-III, we can see that less than five batches are needed for the models to converge to the final F-Score. In the case of MACHO, with only 20 batches both models get very close to their final F-Score. This result confirms what we obtained in Section \ref{subsec:incremental_feat_results}, where we could see that the incremental features converged to their final value with a lower number of observations. 

\begin{table*}
	\centering
	\caption{Classification F-Score on the MACHO training set.}
	\label{tab:macho_classification_results}
    \begin{tabular}{ l r r r r }
        \hline
         Class & IF + SBM & IF + oRF & IF + RF & FATS + RF \\
        \hline
        Quasar & 0.44 & 0.61 & 0.44 & 0.53 \\
        Be Star & 0.78 & 0.26 & 0.61 & 0.78 \\
        Cepheid & 0.50 & 0.71 & 0.85 & 0.92 \\
        RR Lyrae & 0.86 & 0.88 & 0.92 & 0.97 \\
        Eclipsing Binary & 0.64 & 0.57 & 0.65 & 0.76 \\
        MicroLensing & 0.94 & 0.86 & 0.95 & 0.97 \\
        Long Period Variable & 0.85 & 0.92 & 0.97 & 0.94 \\
        \hline
        Weighted Average & 0.82 & 0.79 & 0.86 & 0.91 \\
        \hline
    \end{tabular}
\end{table*}

\begin{table*}
	\centering
	\caption{Classification F-Score on the OGLE-III training set.}
	\label{tab:ogle_classification_results}
    \begin{tabular}{ l r r r r }
        \hline
         Class & IF + SBM & IF + oRF & IF + RF & FATS + RF \\
        \hline
        Cepheid & 0.57 & 0.56 & 0.69 & 0.96 \\
        RR Lyrae & 0.87 & 0.83 & 0.90 & 0.98 \\
        Eclipsing Binary & 0.80 & 0.81 & 0.83 & 0.98 \\
        Long Period Variable & 0.96 & 0.94 & 0.99 & 0.99 \\
        \hline
        Weighted Average & 0.93 & 0.91 & 0.96 & 0.99 \\
        \hline
    \end{tabular}
\end{table*}

\begin{table*}
	\centering
	\caption{Classification F-Score on the CoRoT training set.}
	\label{tab:corot_classification_results}
    \begin{tabular}{ l r r r r }
        \hline
         Class & IF + SBM & IF + oRF & IF + RF & FATS + RF \\
        \hline
        Cepheid & 0.61 & 0.59 & 0.75 & 0.81 \\
        RR Lyrae & 0.54 & 0.51 & 0.57 & 0.66 \\
        Eclipsing Binary & 0.62 & 0.63 & 0.74 & 0.86 \\
        Long Period Variable & 0.00 & 0.00 & 0.00 & 0.00 \\
        \hline
        Weighted Average & 0.32 & 0.30 & 0.38 & 0.41 \\
        \hline
    \end{tabular}
\end{table*}

%\begin{table}
%	\centering
%	\caption{Computational runtime details for model training.}
%	\label{tab:runtime_classification}
%    \begin{tabular}{ l l l }
%        \hline
%         & MACHO & OGLE-III \\
%        \hline
%        SBM & \textbf{7 seconds} & \textbf{6 minutes} \\
%        RF & 11 seconds & 14 minutes \\
%        \hline
%    \end{tabular}
%\end{table} 

\section{Conclusions}
\label{sec:conclusions}

In this work, we present a novel end-to-end streaming model for variable stars classification. Our streaming feature extraction together with the streaming classification model constitutes a significant step towards an efficient real-time classification pipeline to analyze upcoming surveys data. These types of pipelines are a must in scenarios where early detection of alerts are mandatory, such as the upcoming LSST brokers \citep{borne2007}.

Our results show that the incremental features have competitive performance compared to the traditional features while being an order of magnitude faster in a streaming setting. We show how the incremental features get close to their real value with few observations. For example, the incremental period needs around $20\%$ of the light curve observations to converge to the value we get with $100\%$ of the observations. We also can see this in the classification results, where the F-Score stabilizes with few observations. Meanwhile, our classification model performs as good and sometimes better than the streaming Random Forest. As future work, it would be interesting to modify the model in order to be semi-supervised, since in a streaming setting the labels are not always known. We hope this work can help the research community to improve the efficiency in the discovery of new objects of interest in future surveys. We have a Python implementation and it is openly available at \url{https://github.com/lezorich/variational-gmm}. 

\section*{Acknowledgements}

We acknowledge the support from CONICYT-Chile, through the FONDECYT Regular project number 1180054.

%%%%%%%%%%%%%%%%%%%%%%%%%%%%%%%%%%%%%%%%%%%%%%%%%%



\bibliographystyle{mnras}
\bibliography{bibliography} % if your bibtex file is called example.bib

%\begin{thebibliography}{99}
%\end{thebibliography}

%%%%%%%%%%%%%%%%%%%%%%%%%%%%%%%%%%%%%%%%%%%%%%%%%%

%%%%%%%%%%%%%%%%% APPENDICES %%%%%%%%%%%%%%%%%%%%%

%%%%%%%%%%%%%%%%%%%%%%%%%%%%%%%%%%%%%%%%%%%%%%%%%%

% Don't change these lines
\bsp	% typesetting comment
\label{lastpage}
\end{document}